\documentclass[letterpaper, 10 pt, conference]{ieeeconf}  

\IEEEoverridecommandlockouts   

\usepackage{blindtext,graphicx,booktabs,amsmath}
\usepackage{graphicx,amsmath,cite}
\usepackage{wrapfig}
\usepackage{lipsum}
\usepackage{textcomp}
\usepackage{mathtools}
\usepackage{fancyhdr}
\usepackage{graphics}
\usepackage{setspace}
\usepackage{epstopdf}
\usepackage{multirow}
\usepackage{setspace}
\usepackage{booktabs}
\usepackage{array}
\usepackage{amssymb}
\usepackage{verbatim}
\usepackage{cleveref}
\usepackage{color}
\usepackage{relsize}
\usepackage{caption}
\usepackage{subfigure}

\fancypagestyle{firstpage}{\lhead{\vspace{-0.75 cm} \small \textbf{{\fontfamily{cmss}\selectfont
The 3rd IEEE Conference on Control Technology and Applications (CCTA)\\August 19--21, 2019, Hong Kong, China}}}}
\title{\LARGE \bf
Integrated Optimization of Power Split, Engine Thermal Management, and Cabin Heating for Hybrid Electric Vehicles
}

\author{Xun Gong$^{1}$, Hao Wang$^{1}$, Mohammad Reza Amini$^{1}$, Ilya Kolmanovsky$^{2}$, and Jing Sun$^{1}$
\thanks{*This work is supported by the United States Department of Energy (DOE), ARPA-E NEXTCAR programm under award NO. DE-AR0000797}
\thanks{$^{1}$X. Gong, H. Wang, M.R. Amini and J. Sun are with Department of Naval Architecture \& Marine Engineering, University of Michigan, Ann Arbor, MI, 48109, USA. Emails: \{{\tt\small gongxun, autowang, mamini, jingsun@umich.edu}\}}
\thanks{$^{2}$ I. Kolmanovsky is with the Department of Aerospace Engineering, University of Michigan, Ann Arbor, MI, 48109, USA. Email: {\tt\small ilya@umich.edu}} 
}

\begin{document}

\maketitle
\thispagestyle{firstpage}
\pagestyle{empty}

\begin{abstract}
Cabin heating demand and engine efficiency degradation in cold weather lead to considerable increase in fuel consumption of hybrid electric vehicles (HEVs), especially in congested traffic conditions. 
This paper presents an integrated power and thermal management (i-PTM) scheme for the optimization of power split, engine thermal management, and cabin heating of HEVs. 
A control-oriented model of a power split HEV, including power and thermal loops, is developed and experimentally validated against data collected from a 2017 Toyota Prius HEV. Based on this model, the dynamic programming (DP) technique is adopted to derive a bench-mark for minimal fuel consumption, using 2-dimensional (power split and engine thermal management) and 3-dimensional (power split, engine thermal management, and cabin heating) formulations. 
Simulation results for a real-world congested driving cycle show that the engine thermal effect and the cabin heating requirement can significantly influence the optimal behavior for the power management, and substantial potential on fuel saving can be achieved by the i-PTM optimization as compared to conventional power and thermal management strategies. 
\end{abstract}

\vspace{-0.15cm}
\section{Introduction}\label{Secintion:introduction} 
\vspace{-0.15cm}
Concerns of environmental impact and tightened fuel economy regulations  
have motivated numerous technical innovations for vehicles efficiency improvement,
along with the increasing penetration of hybrid electric vehicles (HEVs) and plug-in hybrid electric vehicles (PHEVs).
The HEV power management has been studied extensively and several techniques, from rule-based/heuristic approaches to optimization-based approaches, have been investigated \cite{Sciarretta2007, Silvas2017, Sun2015, MarinaMartinez2016, haowang2018}. 

The previous studies have proven that substantial fuel economy gain can be achieved via advanced power split control methods with the assumption that the combustion engine is operating at its nominal/ideal thermal condition. However, in cold weather, when the vehicle stops or the vehicle operates in the electric drive mode with engine off, the cabin heating requirements may cause fast engine cool-down, 
leading to the loss of energy efficiency or frequent engine start-stop. 
Although the impacts of the thermal effects on the fuel economy have been studied, the integrated power and thermal management in cold weather conditions is still an open topic. In~\cite{Kessels08}, an integrated powertrain control was proposed for an HEV equipped with a natural gas engine, in which the optimization of catalyst light-off time and fuel economy was considered. In~\cite{Arsie10}, an off-line nonlinear constrained optimization was investigated for a specific series hybrid solar vehicle considering engine thermal effects. In~\cite{Gissing2016,Shams-Zahraei2012, Engbroks2018}, globally optimized energy management solutions were investigated for PHEVs equipped with electric heaters using multiple resources for meeting the heating requirement, and the energy saving potentials were illustrated by comparing with conventional strategies. Due to the difference in configurations between PHEVs and HEVs, the results may not be generalized to HEVs.    

In this paper, we propose an integrated power and thermal management (i-PTM) optimization framework for a power split HEV, accounting for cabin heating requirement. The engine is assumed to be the only resource to provide the heating power to the cabin. We first develop and experimentally validate a control-oriented model, including both power and thermal loops.
To provide a bench-mark performance target, the DP is adopted to minimize the fuel consumption by controlling the engine operating mode, power split, and heating power supplied to the cabin, while enforcing the system constraints and managing thermal responses. To demonstrate the thermal impact on behaviors of the power management and the corresponding fuel saving potential, a realistic winter congested driving scenario is considered in the simulation case study in which the proposed i-PTM with 2-dimensional and 3-dimensional DP formulations are compared with the baseline controllers. 


The rest of the paper is organized as follows. Section~\ref{ACC2019:section:modeling} introduces the model and Section~\ref{ACC2019:section:DP} describes the problem formulations and optimization approach for i-PTM system. Section~\ref{ACC2019:section:result} presents the simulation results and fuel saving potentials.
Section~\ref{ACC2019:section:conclusion} summaries the main findings.

\vspace{-0.25cm}
\section{Control-oriented i-PTM system modeling }\label{ACC2019:section:modeling}
\begin{figure}[htp!]
\renewcommand{\captionfont}{\small}
\centering\includegraphics[width=1\linewidth]{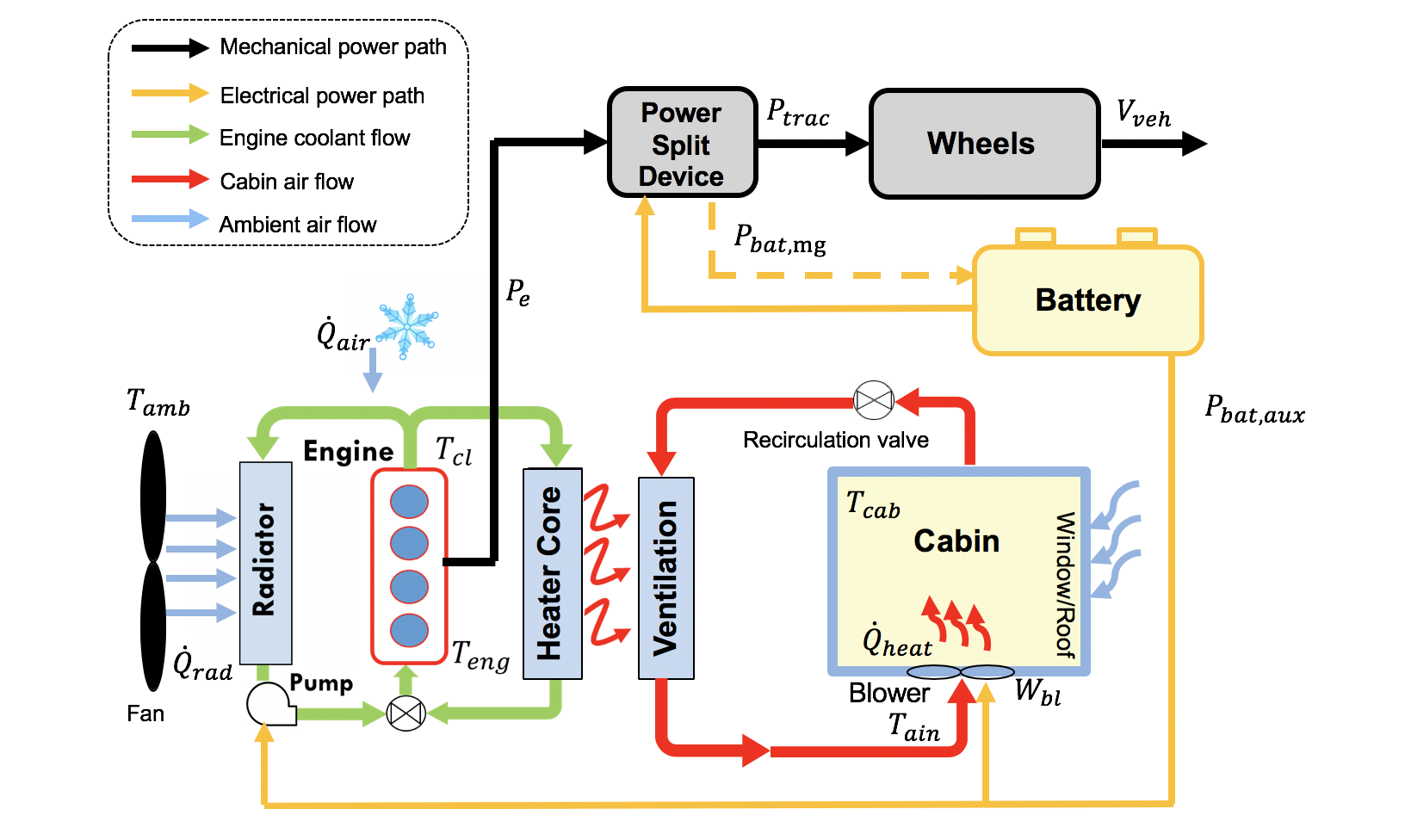}
\caption{Schematic of a power split HEV thermal and power loops for heating scenario.}\vspace{-0.15cm}
\label{fig:HEVmodel:batterydiagram}
\end{figure}

\vspace{-0.15cm}
In this section, a physics-based HEV model is described according to the power split configuration of Toyota Hybrid System (THS \cite{LiuJinm2008,Kim2014}), including the power and thermal loops, and validated against the experimental data collected from our 2017 Prius test vehicle. The overall schematic of the power split HEV for heating scenario is shown in Fig.~\ref{fig:HEVmodel:batterydiagram}. In the power loop, the battery provides the electric power for vehicle traction ($P_{bat,mg}$) and auxiliary devices ($P_{bat,aux}$). The total demanded traction power ($P_{trac}$) is provided via a power split device (PSD) that blends the engine output power ($P_e$) and electric motor traction power. In the thermal loop, considering the cabin heating requirement in winter, the engine is assumed to be the only resource to provide heat power ($\dot{Q}_{heat}$) to the cabin through the heat core. 
Another circulation loop via front radiator/fan will be activated to remove heat ($\dot{Q}_{rad}$) from the engine only when the coolant temperature ($T_{cl}$) is higher than a specific threshold value. In cold weather conditions, the heat loss via air convection ($\dot{Q}_{air}$) is significant in engine thermal loop. 

\vspace{-0.15cm}

\vspace{-0.15cm}
\subsection{Battery Model}
The governing equation of the battery state of charge (SOC) is given by\vspace{-0.15cm}
\begin{equation}
\label{eq:HEVmodel:SOC}
\dot{SOC}=-\frac{U_{oc}-\sqrt{U_{oc}^2-4R_{int} P_{bat}}}{2R_{int}C_{bat}},
\end{equation}
where $P_{bat}$ is the battery power, $C_{bat}$ is the battery capacity. The open-circuit voltage $U_{oc}$ and the internal resistance $R_{int}$ are functions of the battery SOC which are calibrated based on experimental data.
The power provided by the battery is given by\vspace{-0.15cm}
\begin{equation}
\label{eq:HEVmodel:Pbatt}
P_{bat}=P_{bat,mg} + P_{bat,aux},
\end{equation}
where $P_{bat,mg}$ and $P_{bat,aux}$ are the electric power for traction auxiliary devices, respectively. 

\vspace{-0.15cm}
\subsection{Engine Thermal Transients Model}
In this work, the engine coolant temperature represents of the overall engine thermal state variable and its dynamics can be expressed by
\begin{equation}
\label{eq:HEVmodel:dotTcl}
\dot{T}_{cl}=\frac{1}{M_{eng}C_{eng}}(\dot{Q}_{fuel}-P_{e}-\dot{Q}_{exh}-\dot{Q}_{air}-\dot{Q}_{rad}-\dot{Q}_{heat}),
\end{equation}
where $M_{eng}$, $C_{eng}$ and $T_{cl}$ are the mass, the equivalent specific heat capacity, and the coolant temperature of the engine respectively, $\dot{Q}_{fuel}$ is the heat released in the combustion of the fuel, $P_{e}$ is the engine mechanical output power, $\dot{Q}_{exh}$, $\dot{Q}_{air}$ and $\dot{Q}_{rad}$ are the heat rejected by exhaust gas via the convection from the engine to the air and via radiator/fan respectively, $\dot{Q}_{heat}$ is the heat power delivered to the cabin.

\begin{figure}[htp!]
\renewcommand{\captionfont}{\small}
\centering\includegraphics[width=0.95\linewidth]{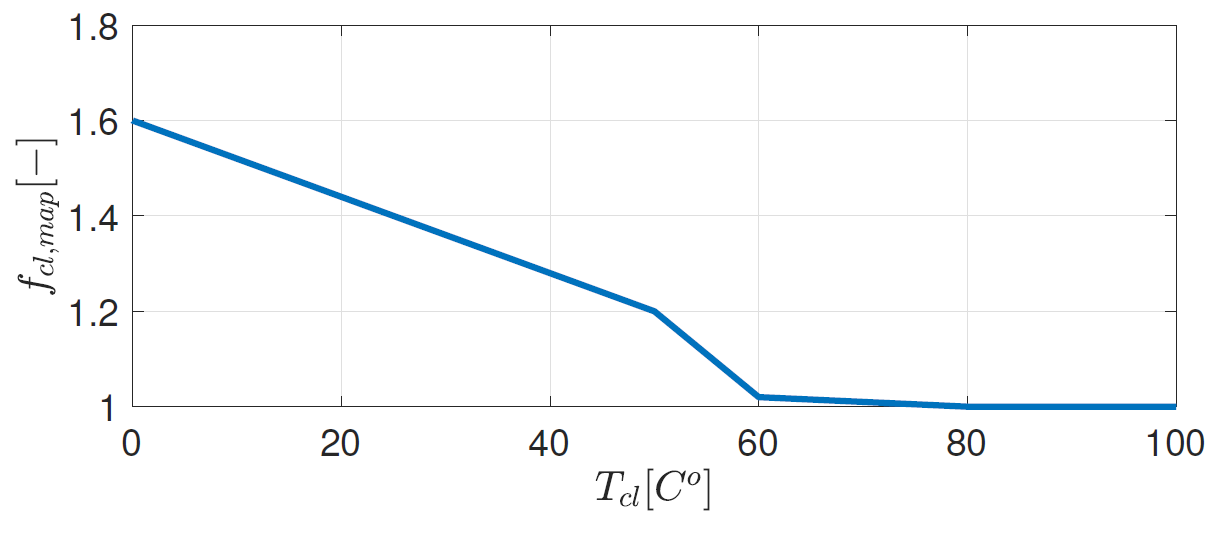}\vspace{-0.25cm}
\caption{Correction factor on fuel consumption reflecting the coolant temperature sensitivity.}\vspace{-0.25cm}
\label{fig:HEVmodel:coolantsensitivity}
\end{figure}

The heat released in combustion is calculated by:
\begin{equation}
\label{eq:HEVmodel:Qfuel}
\dot{Q}_{fuel}=LHV \cdot W_f(\omega_e, \tau_e, T_{cl}),
\end{equation}
where $LHV$ is the low heating value of the gasoline, $W_f$ is the fuel rate calculated by:\vspace{-0.15cm}
\begin{equation}
\label{eq:HEVmodel:fuelconsumption}
W_f = f_{fuel,map} (\omega_e, \tau_e) \cdot f_{cl,map} (T_{cl}),
\end{equation}
where $f_{fuel,map}$ denotes the nominal fuel consumption characterized by a map with engine speed $\omega_e$ and torque $\tau_e$ as the inputs, and $f_{cl,map} (T_{cl})$ is the correction factor reflecting the cold coolant temperature impact on the fuel consumption map. The look-up table adopted from Autonomie simulation software library and calibrated by Argonne National Laboratory~\cite{Kim2014} is shown in Fig.~\ref{fig:HEVmodel:coolantsensitivity}. \vspace{-0.25cm}

The heat rejected by engine exhaust gas is defined as 
\begin{equation}
\label{eq:HEVmodel:Qexh}
\dot{Q}_{exh}=\gamma_{exh}(\dot{Q}_{fuel}-P_{e}),
\end{equation}
where $\gamma_{exh}$ is the exhaust heat coefficient.


The heat rejected by the air convection can be calculated by
\begin{equation}
\label{eq:HEVmodel:Qair}
\dot{Q}_{air}=(T_{cl}-T_{com})A_{eng}\alpha_{eng},
\end{equation}
where $T_{com}$ is the temperature of the engine compartment, $A_{eng}$ is the equivalent heat transfer area and $\alpha_{eng}$ is the specific heat transfer coefficient.
As a simplification, the engine compartment temperature is approximately expressed by a static equation associated with ambient temperature $T_{amb}$ and engine coolant temperature $T_{cl}$.

The heat removed by the radiator/fan is calculated by 
\begin{equation}
\label{eq:HEVmodel:Qrad}
\dot{Q}_{rad} = f_{map,rad}(T_{cl}),
\end{equation}
where the $f_{map,rad}$ is a map calibrated based on the simulation and testing data.

The heating power $\dot{Q}_{heat}$ delivered to cabin is modeled based on the vehicle test data described by the following equation: \vspace{-0.15cm}
\begin{equation}
\label{Task4_Qheat}
\begin{aligned}
\dot{Q}_{heat}  = f(T_{ain},T_{cab,sp},W_{bl}),
\end{aligned}
\end{equation}
where $T_{ain}$ is the vent air temperature,  $T_{cab,sp}$ is the cabin temperature set-point and $W_{bl}$ is air flow through the cabin blower.

\begin{figure}[htp]
\renewcommand{\captionfont}{\small}
\centering\includegraphics[width=0.95\linewidth]
{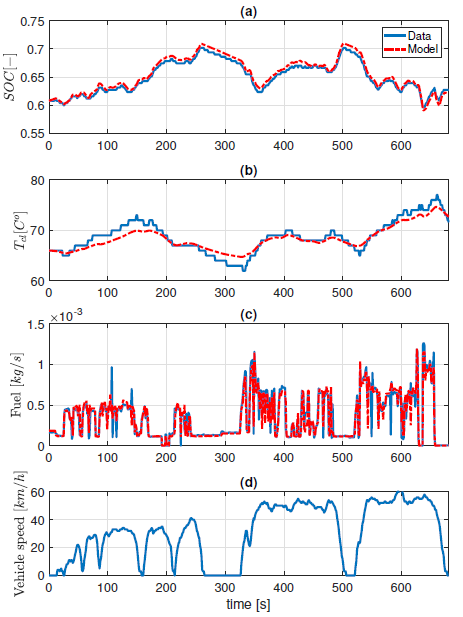}\vspace{-0.25cm}
\caption{Model validation on battery SOC, engine coolant temperature, and fuel flow rate (city driving cycle in Ann Arbor, MI).}\vspace{-0.45cm}
\label{fig:modelvalidation:city}
\end{figure}

The developed battery SOC and engine coolant models are validated against the experimental data collected by driving the HEV test vehicle over real-world city driving cycle in Ann Arbor, Michigan, in January as shown in Fig.~\ref{fig:modelvalidation:city}.
It can be seen that the model ((\ref{eq:HEVmodel:SOC}) and (\ref{eq:HEVmodel:dotTcl})) captures the dynamics well.
\subsection{Cabin Thermal Model}
To evaluate the cabin temperature changes and driver comfort, a cabin thermal model is needed. A simplified cabin average temperature model is considered as follows, 
\begin{equation}
\label{eq:HEVmodel:dotTcab}
\dot{T}_{cab}=\frac{1}{M_{cab}C_{cab}}(\dot{Q}_{heat}+\dot{Q}_{load}+\dot{Q}_{sun}),
\end{equation}
where $M_{cab}$ and $C_{cab}$ represent the equivalent air mass and heat capacity in the cabin respectively, $\dot{Q}_{sun}$ represents the radiation heat from the sun and $\dot{Q}_{load}$ represents the heat load by heat transmission/convection.
All the parameters of the cabin temperature model are obtained from Autonomie software. \vspace{-0.35cm}
\vspace{-0.25cm}
\section{Dynamic Optimization For i-PTM}\label{ACC2019:section:DP}
\vspace{-0.15cm}
The objective of the i-PTM is to obtain the optimal fuel economy while taking the engine thermal condition and cabin heating requirements into account. The i-PTM problem is a multi-state nonlinear constrained optimization problem whose decision variables include integer for engine mode selection, as well as continuous variables.
DP can be adopted to find the optimal solution, but it is computationally demanding and it depends on a given driving cycle. Nevertheless, it provides a benchmark and useful insight to develop online strategies. Thus, DP forms a good framework for solving the i-PTM problem at this stage. \vspace{-0.15cm}
\subsection{Problem Formulation}
\vspace{-0.15cm}
\label{CCTA:sec:Formulation}
In this paper, the fuel saving potential of i-PTM formulations involving different thermal states are discussed. The discretized HEV model described in Section~\ref{ACC2019:section:modeling} can be expressed in a general form as\vspace{-0.15cm}
\begin{equation}
\label{ACC2019:sectionDP:costtogo_N}
x_{k+1} = f(x_k, u_k),
\end{equation}
where $u_k \in {\mathcal{R}}^l$ denotes the vector of control variables and $x_k\in {\mathcal{R}}^n$ denotes the vector of state variables.

\subsubsection{1-dimensional baseline DP} In a conventional power management problem, battery SOC is the only state while battery power $P_{bat}$ and engine operation mode~$e_{mode}$ are selected as control variables~\cite{Sciarretta2007}. This one-state DP addresses fuel economy optimization without explicit consideration of engine thermal condition and the coolant temperature sensitivity on the fuel consumption ($f_{cl,map} \equiv 1$). The 1-dimensional DP is considered as the ``Baseline-DP" for comparison in the simulation study in Section~\ref{ACC2019:section:result}.

\subsubsection{2-dimensional DP} For i-PTM problem incorporating the engine thermal condition, we select the battery SOC ($SOC_k$) and engine coolant temperature ($T_{cl,k}$) as the state variables and the engine operation mode ($e_{mode,k}$), battery power ($P_{bat,k}$) as control variables. We refer to this two-state optimization problem as ``Thermal-DP" for power split and engine thermal management. 
Assuming that the vehicle speed profile and heating power demand ($\dot{Q}_{heat,k}=\dot{Q}_{heat,d}$) are given as known inputs, the optimization problem is to find the control input, $u_k=[e_{mode,k},P_{bat,k}]$, to minimize the overall fuel consumption of the HEV, while enforcing the state constraints and vehicle operating constraints. Thus, with a given driving cycle being discretized by $N$ sampling instants, the optimal energy efficient operation derived by ``Thermal-DP" can be obtained by minimizing the following cost function:\vspace{-0.15cm}
\begin{equation}
\label{ACC2019:sectionDP:costfunction}
min: J(x_k, u_k, N)= \sum_{k=0}^N W_{f,k}(x_k,u_k) + \Phi(x_N),
\end{equation}
where 
$\Phi(x_N)$ is the terminal cost on the states $x_N=[SOC_N, T_{cl,N}]$. 
The constraints on control variables and state variables are imposed by:
\begin{equation}
\label{ACC2019:sectionDP:constraints}
\begin{aligned}
SOC_{min} \leq SOC_k \leq SOC_{max},\\
T_{cl,min} \leq T_{cl,k} \leq T_{cl,max},\\
P_{bat,min} \leq P_{bat,k} \leq P_{bat,max}.\\
\end{aligned}
\end{equation}
Moreover, the vehicle should satisfy the operation constraints according to different operation modes of the engine $e_{mode} \in \{1,2,3\}$, i.e,
(i) Engine off: $e_{mode}=1$, $P_{e,k}=0, \omega_{e,k}=0, W_{f,k}=0$;
(ii) Engine idling: $e_{mode}=2$, $P_{e,k} = 0, \omega_e \neq 0, W_{f,k}=W_{idle}$, where $W_{idle}$ is the fuel rate in idling mode;
(iii) Engine on: $e_{mode}=3$, $P_{e,k} > 0, \omega_{e,k} \neq 0, W_{f,k}=f_{fuel,map} (\omega_{e,k}, T_{e,k}) \cdot f_{cl,map} (T_{cl,k})$.
It should be noted that when $e_{mode}=3$, to reduce the computational load of DP, we assume that (I) the engine operates on the optimal operation line which has been extracted from the experimental data, and (II) the fuel consumption is corrected to reflect the effect of cold temperatures. 

\subsubsection{3-dimensional DP} 
Next, we explore the fuel saving potential via cabin temperature management by coordinating the heating power supply to the cabin. In such a case, we assume that the given heating power demand can be relaxed as long as the cabin temperature is maintained within the desired range.
The cost function~(\ref{ACC2019:sectionDP:costfunction}) can be further extended with the state of the cabin temperature ($T_{cab,k}$) and the additional control variable of heating power ($\dot{Q}_{heat,k}$). The control input $\dot{Q}_{heat,k}$ can be converted to physical control variables based on model~(\ref{Task4_Qheat}). We refer to this three-state optimization problem as ``Thermal-Cabin-DP" for power split, engine thermal management, as well as cabin heating. The corresponding cost function is as follows: \vspace{-0.15cm}
\begin{equation}
\label{ACC2019:sectionDP:costfunction_ex}
min: J(x^{cab}_k, u^{cab}_k, N)= \sum_{k=0}^N W_{f,k}(x^{cab}_k,u^{cab}_k) + \Phi(x^{cab}_N),
\end{equation}
where $x^{cab}_k=[SOC_k, T_{cl,k},T_{cab,k}]$ is the expanded state vector and $u^{cab}_k=[e_{mode,k},P_{bat,k},\dot{Q}_{heat,k}]$ is the expanded control vector. Additional constraints on cabin temperature and heating power are also considered:
\begin{equation}
\label{ACC2019:sectionDP:constraintsEXTR}
\begin{aligned}
T^{LB}_{cab} \leq T_{cab,k} \leq T^{UB}_{cab},\\
\dot{Q}_{heat,min} \leq \dot{Q}_{heat,k} \leq \dot{Q}_{heat,max},\\
\end{aligned}
\end{equation}
where $T^{LB}_{cab}$ and $T^{UB}_{cab}$ denote the predefined lower and upper boundary of the cabin temperature. 

The different DP formulations mentioned above, in terms of state and control variables, are summarized in Table~\ref{Result:table_method}. The corresponding simulation comparison results and the fuel saving potentials will be discussed in Section~\ref{ACC2019:section:result}. \vspace{-0.05cm}
\begin{table}[h!]\vspace{-0.25cm}
\centering
\caption{DP Formulations for i-PTM}
\label{Result:table_method}
\begin{tabular}{|c|c|c|c|}
\hline
 Controller & State & Control input\\ \hline
 \small Baseline-DP             & \small $SOC$         & \small $e_{mode}, P_{bat}$           \\ \hline
\small Thermal-DP   
&\small $SOC, T_{cl}$         & \small $e_{mode}, P_{bat}$ \\ \hline
\small Thermal-Cabin-DP              & \small $SOC, T_{cl}, T_{cab}$         & \small $e_{mode}, P_{bat}, \dot{Q}_{heat}$       \\ \hline
\end{tabular}
\end{table}
\vspace{-0.45cm}
\subsection{Quantization Effects}
DP problems of Section~\ref{CCTA:sec:Formulation} are solved numerically using quantization and interpolation with the sampling time for model~(\ref{ACC2019:sectionDP:costtogo_N}) chosen as $T_s=1 sec$. Considering the ``curse of dimensionality" associated with DP, the grid size should be selected carefully. Since the 3-dimensional ``Thermal-Cabin-DP" becomes intractable with small grids, we limit the grid size to a certain level as shown in Table~\ref{Result:table_quati} to balance the computational load and numerical accuracy. 
\begin{table}[h!]\vspace{-0.25cm}
\centering
\caption{Quantization of state and control variables}
\label{Result:table_quati}
\begin{tabular}{|c|c|c|}
\hline
 Variables & Grids\\ \hline
 \small $SOC$             & \small $SOC_{min}$~:~$0.01$~:~$SOC_{max}$              \\ \hline
\small $T_{cl}$  
&\small $T_{cl,min}$~:~$1~C^o$~:~$T_{cl,max}$          \\ \hline
\small $T_{cab}$              & \small $T^{LB}_{cab}$~:~$1~ C^o$~:~$T^{UB}_{cab}$     \\ \hline
 \small $P_{bat}$             & \small $P_{bat,min}$~:~$0.5~kW$~:~$P_{bat,max}$              \\ \hline
\small $e_{mode}$  
&\small $1:1:3$          \\ \hline
\small $\dot{Q}_{heat}$              & \small $\dot{Q}_{heat,min}$~:~$0.05~kW$~:~$\dot{Q}_{heat,max}$     \\ \hline
\end{tabular}
\end{table}


\section{DP Simulation Results and Discussion}\label{ACC2019:section:result}\vspace{-0.25cm}
\subsection{Real-World Congested Driving Cycle}

\begin{figure}[htp]
\renewcommand{\captionfont}{\small}
\centering\includegraphics[width=0.9\linewidth]
{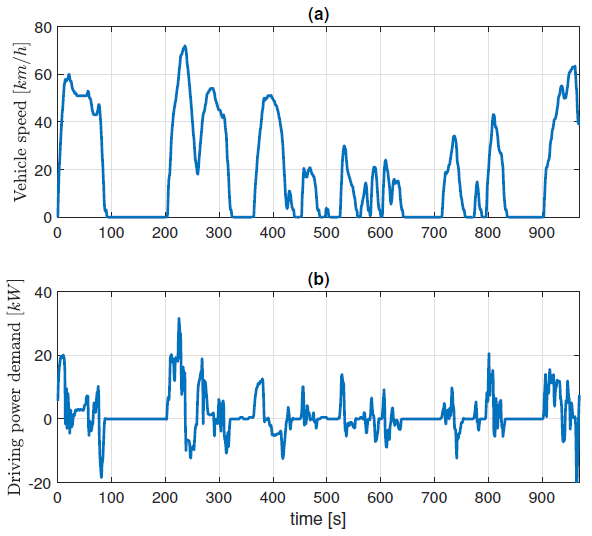}\vspace{-0.25cm}
\caption{The profile of the real-world congested city driving cycle: (a) vehicle speed, (b) driving power demand.}\vspace{-0.45cm}
\label{fig:drivingprofile}
\end{figure}

The simulation is conducted in MATLAB/Simulink environment with the model as described in Section~\ref{ACC2019:section:modeling}. 
In this work, we focus on the congested city driving cycle. To construct the realistic congested driving profile, 
a representative real-world congested driving cycle (16 mins) is recorded and extracted in Ann Arbor, Michigan.
Vehicle speed and demanded traction power are shown in Fig.~\ref{fig:drivingprofile}. 
The ambient temperature of $T_{amb}=-10 C^{o}$ is assumed, which is common during the winter days in Ann Arbor. \vspace{-0.25cm}
\subsection{DP Optimization Results}
\subsubsection{Thermal DP v.s Rule-based Controller} 
\begin{figure}[htp]
\renewcommand{\captionfont}{\small}
\centering\includegraphics[width=1\linewidth]
{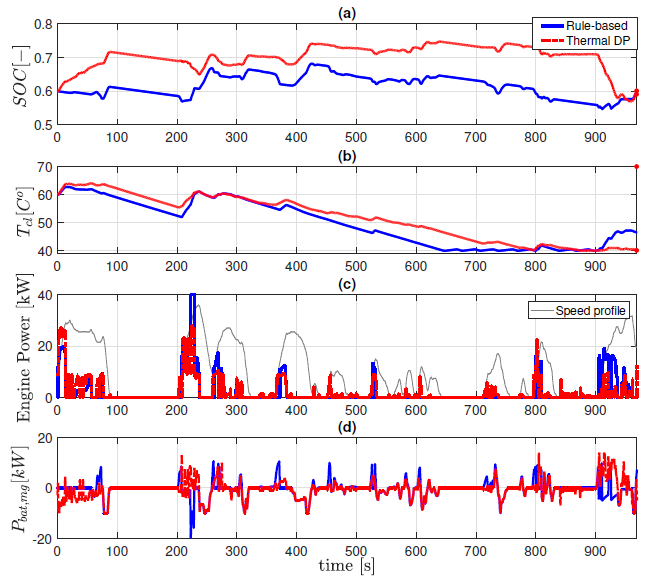}\vspace{-0.25cm}
\caption{Comparison results of Rule-based controller v.s Thermal-DP: (a) battery SOC, (b) engine coolant temperature, (c) engine power (d) battery traction power. $\dot{Q}_{heat}=1500 W$.}\vspace{-0.55cm}
\label{fig:DPresult:ThermalvsRule_1}
\end{figure}

\begin{figure}[htp]
\renewcommand{\captionfont}{\small}
\centering\includegraphics[width=1\linewidth]
{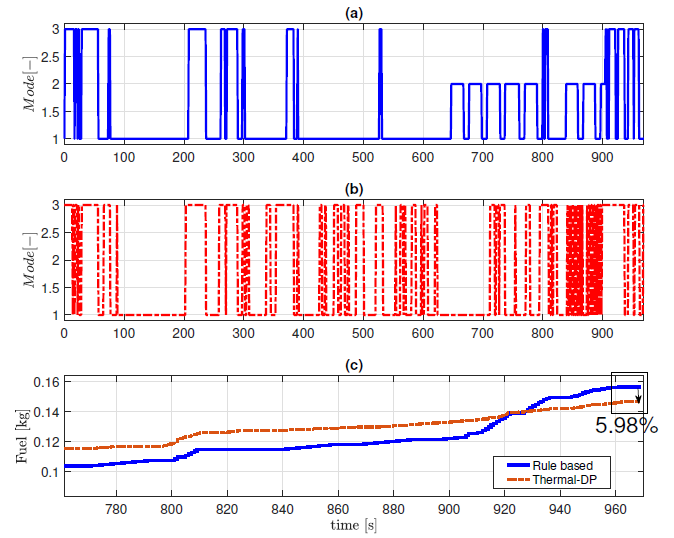}\vspace{-0.25cm}
\caption{Comparison results of Rule-based controller v.s Thermal-DP: (a) engine mode of rule-based controller, (b) engine mode of thermal-DP, (c) fuel consumption. $\dot{Q}_{heat}=1500 W$.}\vspace{-0.55cm}
\label{fig:DPresult:ThermalvsRule_2}
\end{figure}

We first compare the proposed ``Thermal-DP" with a rule-based controller that is based on a load leveling logic with well tuned parameters, see~\cite{Liujinm2005}. In this case, we assume the heating power is given as a constant $\dot{Q}_{heat}=1500 W$. 
To prevent the engine temperature from dropping too low, the rule-based controller will start the engine with idling status even there is no traction power requirement when $T_{cl} < 40 C^{o}$~\cite{Tashiro2016}.
Compared with rule-based controller, ``Thermal DP" uses the engine more often but in a relatively lower load level as shown in Fig.~\ref{fig:DPresult:ThermalvsRule_1} (c) and Fig.~\ref{fig:DPresult:ThermalvsRule_2} (a) which leads to higher coolant temperature and SOC. As the coolant temperature drops down, the rule-based controller turns on the engine with idle mode to enforce thermal constraint, while the ``Thermal-DP" suggests to start the engine in charging mode at vehicle stops (820 s - 900 s) to keep the coolant temperature warm while charging the battery. 
The fuel saving by DP is nearly 6\%. 
\subsubsection{Baseline DP v.s Thermal DP}
\begin{figure}[htp!]
\renewcommand{\captionfont}{\small}
\centering\includegraphics[width=0.95\linewidth]
{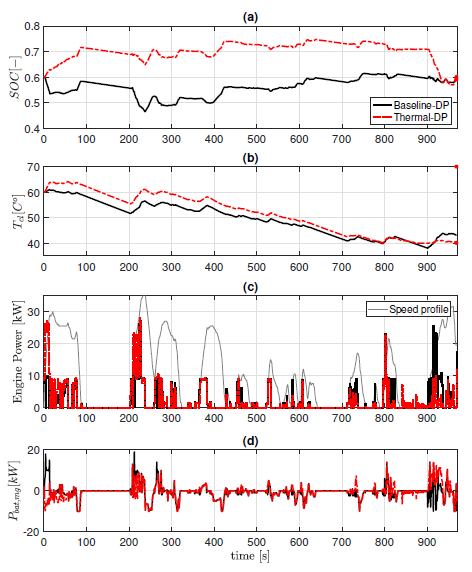}\vspace{-0.25cm}
\caption{Comparison results of Baseline DP v.s Thermal-DP: (a) battery SOC, (b) engine coolant temperature, (c) engine power (d) battery traction power. $\dot{Q}_{heat}=1500 W$.}\vspace{-0.25cm}
\label{fig:DPresult:ThermalvsBase_1}
\end{figure}

\begin{figure}[htp!]
\renewcommand{\captionfont}{\small}
\centering\includegraphics[width=0.95\linewidth]
{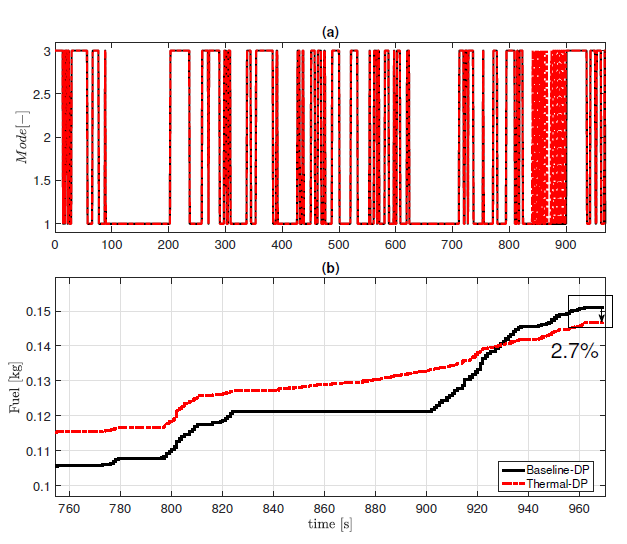}\vspace{-0.25cm}
\caption{Comparison results of Baseline DP v.s Thermal-DP: (a) engine mode, (b) fuel consumption. $\dot{Q}_{heat}=1500 W$.}\vspace{-0.25cm}
\label{fig:DPresult:ThermalvsBase_2}
\end{figure}

 In order to further illustrate the impact of engine thermal effect on the optimized power management of the HEV, the results of ``Thermal DP" are compared against the ``Baseline DP" which does not consider the engine coolant temperature as a state and its effect on fuel consumption in optimization.  
 The comparison results are shown in
 Fig.~\ref{fig:DPresult:ThermalvsBase_1}. 
 Compared with the ``Baseline DP", the ``Thermal DP" makes the engine work harder at the beginning of the trip when the engine coolant temperature is relatively high. By doing this, more electric energy is stored into the battery (Fig.~\ref{fig:DPresult:ThermalvsBase_1} (a)) and thermal energy is stored in coolant as reflected by the higher coolant temperature (Fig.~\ref{fig:DPresult:ThermalvsBase_1} (b)). In contrast, towards the end of the trip, the ``Thermal DP" releases the stored electric energy into traction power which avoids much engine operation at high load when the coolant temperature is low. From Fig.~\ref{fig:DPresult:ThermalvsBase_2} (a), it can be seen that the engine start-stop timing is almost the same,
 however, the power split ratio is totally different. The ``electric and thermal storage concept" leads to $2.7\%$ fuel saving potential.      
\vspace{-0.25cm}
\subsubsection{Thermal-Cabin-DP}
\begin{figure}[htp!]
\renewcommand{\captionfont}{\small}
\centering\includegraphics[width=1\linewidth]
{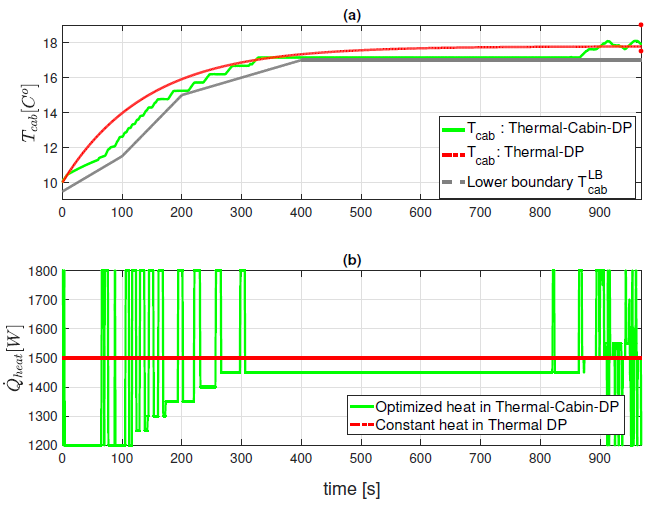}\vspace{-0.25cm}
\caption{Results of Thermal-Cabin-DP: (a) cabin temperature, (b) heating power.}\vspace{-0.35cm}
\label{fig:DPresult:cabin}
\end{figure}

The results of i-PTM optimization which applies the ``Thermal-Cabin-DP"
are presented in Fig.~\ref{fig:DPresult:cabin}. In this case, 
the heating power $\dot{Q}_{heat}$ is allowed to vary in the range of $1200W-1800W$. To manage the cabin the temperature, the terminal value of the cabin temperature $T_{cab,N}$ is set as the same as the case of $\dot{Q}_{heat}=1500W$ and the lower bound of the cabin temperature $T^{LB}_{cab}$ is pre-defined. As shown in Fig.~\ref{fig:DPresult:cabin} (a) and (b), to minimize the fuel consumption, DP suggests the driver to compromise the cabin thermal comfort by manipulating the heating power and make the cabin temperature to follow the predefined lower bound trajectory for further fuel savings. Note that the power split strategy of ``Thermal-Cabin-DP" is almost the same with  ``Thermal-DP", however, by slowing down the warm-up period and slightly reducing the steady-state temperature, the average cabin temperature is reduced by $0.5~C^{o}$ while nearly $1\%$ fuel can be saved. 

\begin{figure}[htp!]
\renewcommand{\captionfont}{\small}
\centering\includegraphics[width=1\linewidth]
{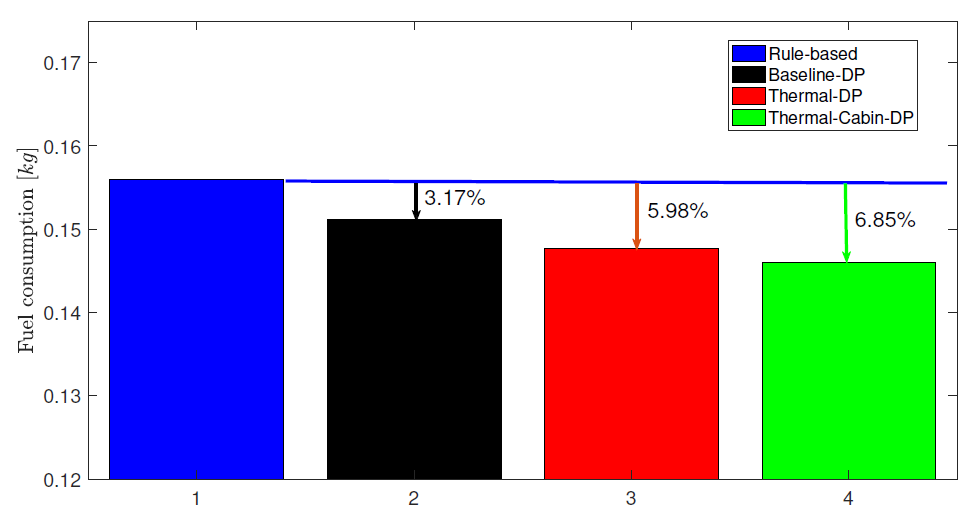}\vspace{-0.25cm}
\caption{Results of fuel saving applying different strategies.}\vspace{-0.35cm}
\label{fig:DPresult:fuelbar}
\end{figure}

Fig.~\ref{fig:DPresult:fuelbar} summaries the overall fuel consumptions of the HEV with different strategies. Up to $6.85\%$ fuel saving can be achieved compared with the rule-based controller. Note that the computation time of the ``Thermal-DP" and ``Thermal-Cabin-DP" are approximate 1 hour and 40 hours respectively on a computer with Intel Core i7 @ 2.6GHz CPU.\vspace{-0.25cm}
%
%


\section{CONCLUSIONS}\label{ACC2019:section:conclusion}\vspace{-0.15cm}
 This paper presents the integrated power and thermal management for a power split HEV during cold winter days in congested city driving scenario. The objective is to optimize the fuel efficiency while accounting for the engine thermal condition and cabin heating requirements. The dynamic programming technique is adopted to find the optimal solution based on an experimentally validated model of the power-split HEV that includes power and thermal loops. The simulation results of a real-world congested driving cycle show that the optimal power management, as well as the fuel consumption, change significantly when the engine thermal condition is incorporated. Up to 6.85\% fuel saving potential is demonstrated by integrated power and thermal optimization compared with a well-calibrated rule-based controller. 

\addtolength{\textheight}{-12cm}   




\bibliographystyle{IEEEtran}
\bibliography{ref}





\end{document}